\documentclass[conference]{IEEEtran}
\IEEEoverridecommandlockouts
\usepackage{cite}
\usepackage{amsmath,amssymb,amsfonts}
\usepackage{algorithmic}
\usepackage{graphicx}
\usepackage{textcomp}
\usepackage{xcolor}
\def\BibTeX{{\rm B\kern-.05em{\sc i\kern-.025em b}\kern-.08em
    T\kern-.1667em\lower.7ex\hbox{E}\kern-.125emX}}

\usepackage{hyperref}
\usepackage[switch]{lineno}

\usepackage{amsmath}
\usepackage{amssymb}
\usepackage{bm}
\usepackage{graphicx}
\usepackage{subfig}
\usepackage{multirow}
\usepackage{booktabs}

\begin{document}

\title{Learning Dynamic Preference Structure Embedding From Temporal Networks
}

\author{
\IEEEauthorblockN{
    Tongya Zheng\textsuperscript{1},
    Zunlei Feng\textsuperscript{2},
    Yu Wang\textsuperscript{2},
    Chengchao Shen\textsuperscript{3},
    Mingli Song\textsuperscript{1,*\thanks{*Corresponding author}},\\
    Xingen Wang\textsuperscript{1},
    Xinyu Wang\textsuperscript{1},
    Chun Chen\textsuperscript{1},
    Hao Xu\textsuperscript{4}
    }
    \IEEEauthorblockA{
        \textsuperscript{1}\textit{College of Computer Science, Zhejiang University},
        \textsuperscript{2}\textit{School of Software Technology, Zhejiang University},\\
        \textit{\{tyzheng,brooksong,newroot,wangxinyu,chenc\}@zju.edu.cn, 
        \{zunleifeng,yu.wang\}@zju.edu.cn}\\
        \textsuperscript{3}\textit{Central South University},\textsuperscript{4}\textit{Zhejiang Lab}\\
        \textit{scc.cs@csu.edu.cn,xuhao.econ@outlook.com}
    }
    \footnote{*Corresponding author}
}

\maketitle

\begin{abstract}
    The dynamics of temporal networks lie in the continuous interactions between nodes, which exhibit the dynamic node preferences with time elapsing.
    The challenges of mining temporal networks are thus two-fold: the dynamic structure of networks and the dynamic node preferences.
    In this paper, we investigate the dynamic graph sampling problem, aiming to capture the preference structure of nodes dynamically in cooperation with GNNs. 
    Our proposed Dynamic Preference Structure (DPS) framework consists of two stages: structure sampling and graph fusion.
    In the first stage, two parameterized samplers are designed to learn the preference structure adaptively with network reconstruction tasks.
    In the second stage, an additional attention layer is designed to fuse two sampled temporal subgraphs of a node, generating temporal node embeddings for downstream tasks.
    Experimental results on many real-life temporal networks show that our DPS outperforms several state-of-the-art methods substantially owing to learning an adaptive preference structure.
    The code will be released soon at https://github.com/doujiang-zheng/Dynamic-Preference-Structure.
\end{abstract}


\section{Introduction}
Networks in real-life scenarios usually present dynamic interactions between nodes, such as online payment networks, movie rating networks, and question-answering networks~\cite{bennett2007netflix,kunegis2013konect,snapnets,network-data}.
Researchers are interested in mining node preferences of networks to make precise predictions for account maliciousness~\cite{kumar2019jodie,wang2020apan}, commodity recommendation~\cite{zuo2018htne,tgat_iclr20}, and so on.
In contrast to static networks, temporal networks exhibit dynamic node preferences through the continuing dynamic interactions.
For example, a user likes watching both action movies and comedy movies in a static movie rating network; however, the user may show distinctive preferences for movie genres at different time points when the movie ratings are labeled with timestamps.

\begin{figure}[t]
    \centering
    \includegraphics[width=.9\linewidth]{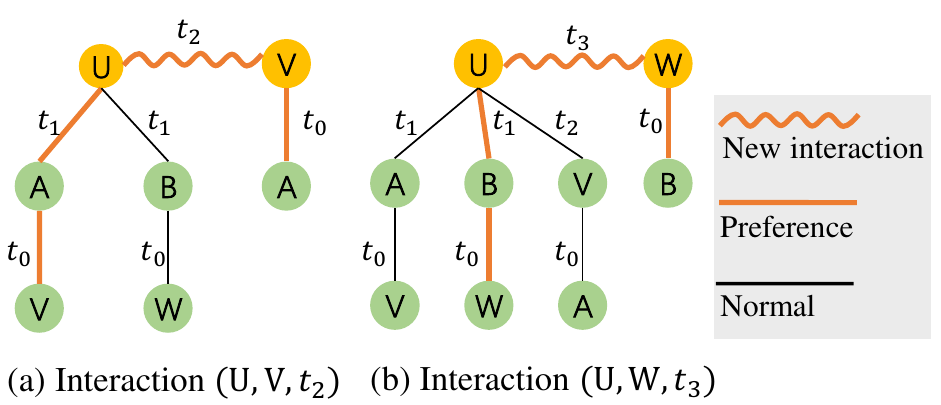}
    \caption{
    The preference structures (highlighted in orange) of node U are dynamic with respect to time, contributing to the new interactions.
    (a) A new interaction (U, V, $t_2$) is related with the previous interaction (U, A, $t_1$) and U's second-order interaction (A, V, $t_0$). 
    (b) Another new interaction (U, W, $t_3$) is related with the previous interaction (U, B, $t_1$) and U's second-order interaction (B, W, $t_0$).}
    \label{fig:example}
\end{figure}

As shown in Fig.~\ref{fig:example}, challenges to graph representation learning on temporal networks are two-fold.
The first challenge is learning the dynamic structure of temporal networks.
Fig.~\ref{fig:example} depicts the expansion of node U's neighbors from $t_2$ to $t_3$, while other nodes except V and W keep the same neighborhood.
The asynchronous interactions among nodes will change the graph topology thousands of times even in a short period, challenging existing methods on a fixed graph topology.
The second challenge is identifying the preference structures of nodes with time elapsing.
Fig.~\ref{fig:example} shows the different preference structures of node U's two new interactions, which obey the common-neighbor assumption.
The interaction (U, V, $t_2$) is attributed to the temporal path U$\overset{t_1}{\rightarrow}$A$\overset{t_0}{\rightarrow}$V, where A is the common neighbor of \{U, V\};
while the interaction (U, W, $t_3$) is attributed to the temporal path U$\overset{t_1}{\rightarrow}$B$\overset{t_0}{\rightarrow}$W, where B is the common neighbor of \{U, W\}.
It requires both mining high-order relations in networks and catching up with dynamic node preferences.

Previous methods mainly resort to recurrent mechanisms to capture the dynamic node preferences, e.g., recurrent neural networks~\cite{kumar2019jodie,tnode,evolvegcn}, and temporal point process~\cite{zhou2018dynamic,zuo2018htne,trivedi2018dyrep}.
The recurrent mechanism uses nodes' historical interactions to predict future behaviors, which pays more attention to sequential modeling than network modeling.
Recently, graph neural networks (GNNs)~\cite{kipf2016semi,hamilton2017graphsage,velivckovic2017gat} have shown great success in graph learning tasks.
Compared with GNNs, the recurrent mechanism loses the high-order collaborative filtering signal in networks~\cite{wang2019ngcf,attention2017vaswani}.
For instance, recurrent methods are not good at inferring the second-order proximity between node U and node V in Fig.~\ref{fig:example}(a).
Although some hybrid methods bridge the gap between sequential modeling and network mining~\cite{tnode,evolvegcn,zhou2018dynamic}, these methods rely on a sequence of static networks and cannot handle the dynamic structure of temporal networks adaptively.
However, recent proposed temporal GNNs~\cite{wang2020apan,tgat_iclr20} based on heuristic sampling strategies can hardly capture the dynamic node preferences.
These heuristic strategies sample nodes' neighborhoods uniformly or chronologically and take no account of personalized preference structures, which may fail in long-range preference modeling.

To overcome the two challenges mentioned above, we investigate the problem of sampling from the dynamic structure of temporal networks adaptively, making GNNs aware of temporal patterns.
Our motivations are three-fold.
First, it is reasonable to generate high-quality node embeddings with a few sampled neighbors~\cite{wang2020apan,tgat_iclr20,hamilton2017graphsage,fastgcn}.
Second, the preference structure of temporal networks will evolve forward with time elapsed, as shown in recurrent methods~\cite{zuo2018htne,tnode,zhou2018dynamic,nguyen2018ctdne}.
Third, sampling methods can significantly boost the training efficiency of GNNs~\cite{hamilton2017graphsage,fastgcn}.

Our proposed Dynamic Preference Structure (DPS) framework consists of two stages: structure sampling and graph fusion.
In the first stage, we devise two parametrized samplers to capture the preference structure adaptively: the \textbf{T}ime \textbf{D}ecay \textbf{S}ampling (TDS) is proposed to capture the temporal patterns of each node with a parametrized time decay distribution; while the \textbf{G}umbel \textbf{A}ttention \textbf{S}ampling (GAS) is proposed to encode the semantic proximity of nodes with a shallow graph neural network.
In the second stage, we devise an attention-based fusion layer to fuse the node embeddings from different sampled subgraphs for each node.
These sampled subgraphs are generated from the dynamic structure of temporal networks by the pre-trained TAS and GAS samplers, respectively.
The temporal link prediction is used as a pretext task to train both stages of the DPS framework.
Finally, experimental results on many real-life temporal networks show that our proposed DPS outperforms several state-of-the-art methods.

Our contributions can be summarized as follows:
\begin{itemize}
    \item We devise two parametrized samplers TDS and GAS, to select the preference structures of temporal networks with network reconstruction tasks.
    \item We further devise an attention-based fusion layer to fuse node embeddings from different temporal subgraphs sampled by TDS and GAS. 
    \item Experiments are conducted over a wide range of temporal networks, and results demonstrate the efficiency of our proposed DPS framework.
\end{itemize}

\section{Related Works}

\subsection{Static Graph Embedding}

Recently, with the widespread use of deep learning techniques in computer vision~\cite{alexnet} and natural language processing~\cite{word2vec}, researchers are inspired to propose several deep graph learning methods~\cite{gnn2020survey,embedding2018survey}, which mainly consist of two categories: skip-gram models~\cite{embedding2018survey} and graph neural networks~\cite{gnn2020survey}.
On the one hand, DeepWalk~\cite{perozzi2014deepwalk} is the pioneering skip-gram model following word2vec~\cite{word2vec}, encoding the random walks of graphs as sampled node sequences.
Node2Vec~\cite{grover2016node2vec} further explores the width and depth of the random walk strategy with two controllable parameters, while LINE~\cite{tang2015line} encodes the high-order proximity between nodes in the graph topology.
On the other hand, researchers~\cite{deffe2016gcn} attempt to define the graph convolution operation in the non-euclidean space from graph signal processing.
GCN~\cite{kipf2016semi} simplifies existing GNNs and applies successfully in semi-supervised node classification.
GAT~\cite{velivckovic2017gat} further introduced a self-attention mechanism to compute the weights of neighbors adaptively and achieves substantial improvements on node classification.


\subsection{Temporal Network Embedding}

Existing researches involve temporal networks from a wide range of scenarios, including citation and collaboration~\cite{zhou2018dynamic,trivedi2018dyrep,zhang2020tigecmn}, commodity purchasing~\cite{zuo2018htne,kumar2019jodie}, fraud detection~\cite{wang2020apan}.
From the property of graphs, previous methods can be divided into two categories: discrete-time graph (DTG) methods and continuous-time graph (CTG) methods.
On the one hand, the input of DTG methods~\cite{zhou2018dynamic,tnode} is a sequence of graph snapshots, where the graph topology in each snapshot will evolve chronologically.
This kind of method is unable to handle edges at the finest granularity (e.g., at a time scale of seconds)~\cite{nguyen2018ctdne}.
On the other hand, CTG methods are designed to generate temporal node embeddings with the upcoming new interactions.
Methods following the recurrent paradigm, like HTNE~\cite{zuo2018htne}, DyRep~\cite{trivedi2018dyrep}, JODIE~\cite{kumar2019jodie}, and TigeCMN~\cite{zhang2020tigecmn}, update node embeddings with new interactions.
Recurrent methods are not yet powerful in learning the graph topology.
Not until recently, TGAT~\cite{tgat_iclr20} and APAN~\cite{wang2020apan} are proposed for inductive learning on temporal networks using graph neural networks.

\section{Method}

\begin{figure*}[t]
    \centering
    \includegraphics[width=.9\textwidth]{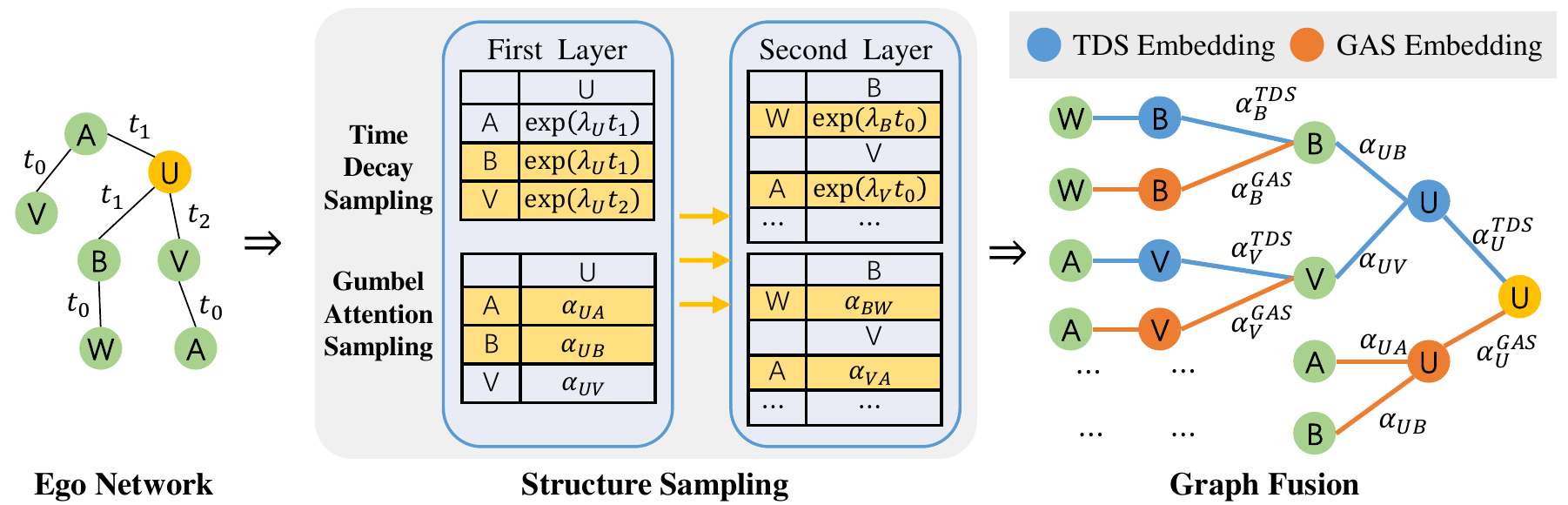}
    \caption{
      The architecture of the 2-layer Dynamic Preference Structure (DPS) framework.
      Given a node U before $t_3$, we extract its 2-hop ego networks recursively.
      In the structure sampling stage, samplers, including Time Decay Sampling (TDS) and Gumbel Attention Sampling (GAS), valuate the neighborhoods for each node and sample two neighbors according to their importance distributions.
      The sampling procedure is repeated two times recursively.
      In the graph fusion stage, an attention-based fusion layer combines node embeddings from different subgraphs, each of which generates a specified node embedding. 
    }
    \label{fig:framework}
\end{figure*}

Figure~\ref{fig:framework} depicts that our proposed \textbf{D}ynamic \textbf{P}reference \textbf{S}tructure (DPS) framework consists of two stages: structure sampling, and graph fusion, aiming at providing temporal node embeddings with respect to time. 
Firstly, we devise Time Decay Sampling (TDS) and Gumbel Attention Sampling (GAS) to capture the preference structures adaptively.
As shown in Fig.~\ref{fig:framework}, TDS samples the neighborhood according to a normalized categorical distribution $p_U = f(N(U);\lambda_U)$, parametrized by a learnable personalized $\lambda_U$ for the node U. 
Meanwhile, GAS selects the most significant neighbors according to the attention coefficients of a one-layer graph attention network. 
Secondly, in the graph fusion stage, each sampled subgraph generates a specified node embedding.
Then a fusion layer is devised to combine the node embeddings from different subgraphs sampled by TDS and GAS.
The temporal link prediction is used as a pretext task to train both stages.

\subsection{Problem Definition}

A temporal network can be defined as $G = (V, E, \mathcal{T})$, where $E$ is a set of temporal edges (interactions), and $\mathcal{T}:E\rightarrow \mathbb{R}^+$ is a function that maps each edge to a corresponding timestamp\cite{nguyen2018ctdne}. 
Our target of graph embedding is to learn an embedding function revealing the underlying edge generation distribution that neighbors of node $u$ at time $t$ can be drawn from $v \sim p(\cdot | u, t)$ in a temporal network~\cite{perozzi2014deepwalk,nguyen2018ctdne,yang2020understanding}.
The temporal node embedding problem can be written as
\begin{equation}
  \underset{h}{\arg\max} \text{ } p_{(u,t)}(v) = \frac{exp (h(v, t) \cdot h(u, t)) }{ \sum_{v' \in V} exp (h(v', t) \cdot h(u, t)) },
  \label{eq:mle}
\end{equation}
where $h(\cdot, t)$ is the embedding function, the dot product is the similarity function, $(u, v, t)$ is an observed interaction, $t$ is the time point, and $v'$ iterates all nodes of $V$.

\subsection{Structure Sampling}
The structure of a temporal network keeps evolving with time elapsing, called the dynamic structure.
Since the graph convolution works on a fixed graph topology~\cite{kipf2016semi,hamilton2017graphsage,velivckovic2017gat}, the structure sampling stage aims at capturing the preference structure of the temporal network with respect to time.
On the one hand, the proposed Time Decay Sampling (TDS) is motivated by the influence decay of interactions with time elapsing.
On the other hand, the proposed Gumbel Attention Sampling (GAS) is motivated by the semantic proximity of nodes by graph neural networks (GNNs)~\cite{tgat_iclr20,kipf2016semi}. 

\subsubsection{Time Decay Sampling}
TDS valuates the neighborhood significance of historical interactions by learning a personalized time-decay distribution for each node $u$.
Let $N(u, t) = \{(u, v, t_k) | (u, v, t_k) \in E \textrm{ and } t_k < t\}$ be the neighborhood interactions of node $u$ before $t$, and $t_k$ be the timestamp of an interaction.

\textbf{Exponential decay.}
Intuitively, a node's historical interactions show reduced influences on its future behaviors with respect to the elapsed time.
The function of TDS samples a bunch of interactions from $N(u, t)$ according to the reserved influences of $N(u, t)$.
The normalized influences of $N(u, t)$ is here a categorical distribution following the exponential decay rate, written as
\begin{equation}
\begin{split}
    p_{(u,t)}(t_k) &= \frac{ \lambda_u exp(-\lambda_u (t - t_k))}{\sum_{w \in N(u,t)} \lambda_u exp(-\lambda_u (t - t_w))}, \\
    &= \frac{exp(\lambda_u t_k)}{\sum_{w \in N(u,t)} exp(\lambda_u t_w )},
    \label{eq:cate-prob}
\end{split}
\end{equation}
which is only related to the temporal proximity. 
Once getting $\lambda_u$ and the corresponding interaction set $N(u, t)$, we sample a bunch of interactions from $N(u, t)$, denoted by
\begin{equation}
    S(u, t) = \{(u, v, t_k) \sim p_{(u,t)}(\cdot | \lambda_u)\}, s.t. \vert S \vert = s,
    \label{eq:tds-sample}
\end{equation}
where $s$ is the sample size.

\textbf{TDS optimization.}
To avoid the burdensome hyper-parameter search over $\lambda_u$, we propose to use temporal link repetition to estimate a proper decay rate $\lambda_u$. 
The task is to find the interaction $(u, v, t)$ that repeated at the latest time $t_v$ in $N(u, t)$.
The probability of the repetition is denoted by $p_{(u,t)}(t_v)$, as defined in Eq.~(\ref{eq:cate-prob}).
The objective for a specified node $u$ is maximizing the likelihood of the temporal link repetition by iterating over the interactions of $u$, written as
\begin{align}
    \underset{\lambda_u \in \mathbb{R}^+}{\arg\max} \text{ } log \underset{(u, v, t) \in E}{\Pi} p_{(u, t)}(t_v),
    \label{eq:tds-convex}
\end{align}
where the decay rate is constrained $\lambda_u  \le 100$ for numerical stability.
It is a well-known log-sum-exp convex problem, which can be tackled by convex programming solvers~\cite{cvxpy,cvxpy2}.
This task can well capture the repeated temporal patterns of node $u$ with $\lambda_u$, indicating the personalized node preference.
Moreover, interactions that didn't happen in the past are ignored in Eq.~(\ref{eq:tds-convex}).
Analysis of those interactions requires high-order methods such as GNNs~\cite{wang2019ngcf}.

\subsubsection{Gumbel Attention Sampling}
TDS is designed to capture personalized node preference of temporal patterns.
However, the lack of latent embedding methods makes it hard to understand the semantic proximity of nodes.
We combine the Gumbel-softmax trick~\cite{gumbel-softmax}, which enables training neural networks with efficient sampling from discrete distributions, and GNNs~\cite{kipf2016semi,hamilton2017graphsage,velivckovic2017gat}, which have achieved remarkable success in graph learning, to capture the semantic preference structure of temporal networks.
Like TDS, the goal of Gumbel Attention Sampling (GAS) is sampling a bunch of interactions from $N(u, t)$ when given a node $u$ and a time point $t$.

\textbf{Gumbel-softmax trick.}
Let $\bm{h}_u^t$ be the temporal node embedding of $(u, t)$ at the input layer generated by the GNN.
Then the unnormalized attention score of $(v, t_k)$ with respect to $(u, t)$ via the interaction $(u, v, t_k)$ is formulated as
\begin{equation}
    p_{(u, t)} (v, t_k) = (W^Q \bm{h}_v^{t_k}) \cdot (W^K \bm{h}_u^t),
\end{equation}
where $W^Q$ is the query transform matrix, $W^K$ is the key transform matrix~\cite{attention2017vaswani}, and $\cdot$ is the inner product between vectors.
Despite the attention scores could represent the semantic proximity between nodes, the attention mechanism is not well suited for sampling the preference structure to provide better temporal node embeddings.
The Gumbel-softmax trick~\cite{gumbel-softmax} is further introduced to enable the training process of sampling from attention scores.
The Gumbel-softmax probability for $(u, v, t_k)$ in $N(u, t)$ is computed as
\begin{equation}
    \begin{split}
    \alpha_{uv} &= \frac{exp((p(v,t_k) + g_v) / \tau)}{\sum_{w \in N(u, t)}exp((p(w, t_w) + g_w) / \tau)}, \\
    g &= -log(-log(\epsilon)), \textrm{where } \epsilon \sim Uniform(0,1),
    \label{eq:gas-attention}
    \end{split}
\end{equation}
where $g$ is drawn from a Gumbel distribution, and $\tau$ is the softmax temperature.
At the beginning of training, a large temperature $\tau$ makes the attention scores more smooth.
During training, the attention scores approaches the real distribution $p_{(u,t)}$ as the temperature $\tau \rightarrow 0$.

\textbf{GAS optimization.}
Temporal link prediction, predicting the probability of nodes' future links, is used to train the one-layer GNN of GAS.
In each forward pass, GAS samples $s$ interactions of the node $u$ according to
\begin{equation}
    S(u, t) = {top}_s\{\alpha_{uv}, \forall (u,v,t_k) \in N(u, t)\}.
    \label{eq:gas-sample}
\end{equation}
The temporal node embedding of $u$ at the first layer is computed as 
\begin{equation}
    h_u^1(t) = \sum_{(v, t_k) \in S(u, t)} \alpha_{uv} W^V \bm{h}_v^{t_k},
\end{equation}
where $W_V$ is the value transform matrix.
The new node embedding is a weighted sum of selected interactions.
The temporal link prediction is used as the training task since it describes the node dynamics detailedly, as the temporal link repetition of Eq.~(\ref{eq:tds-convex}) in TDS optimization.

For simplicity, we adopt the same graph convolution architecture and training objective in Section~\ref{sec:graph-fusion} and Section~\ref{sec:model-opt} for training the GAS sampler.
Section~\ref{sec:graph-fusion} depicts the feature construction of an interaction in Eq.~(\ref{eq:edge-feat}), and the first-order graph attention convolution in Eq.~(\ref{eq:first-order}).
Due to the dynamic structure of temporal networks, GAS only uses a one-layer network for graph convolution.
Section~\ref{sec:model-opt} trains the neural network with the negative sampling technique~\cite{perozzi2014deepwalk,word2vec} to boost the training efficiency.

\subsection{Graph Fusion}
\label{sec:graph-fusion}

\subsubsection{Sampling-based Graph Convolution}
Our proposed TDS and GAS samplers are pretrained according to Eq.~(\ref{eq:tds-convex}) and Eq.~(\ref{eq:loss-ce}) respectively.
Let $S^{TDS}(u, t)$ and $S^{GAS}(u, t)$ be the sampled neighbor set of node $u$ at time $t$.
During graph convolution, these pre-trained samplers replace the heuristic uniform sampler~\cite{tgat_iclr20,hamilton2017graphsage} to retrieve the preference structure for a specified node.

An interaction $e=(u, v, t_k)$ in the neighbor set carries multiple information: the neighbor identity $v$, the timestamp $t_k$, and the interaction feature $\bm{m}_{uv}^{t_k}$.
The information is then transformed into dense vectors by some encoding functions for back-propagation in neural networks.
Let $\bm{h}_v^{l-1}$ be the node embedding of $v$ at $(l-1)$-th layer, which is a one-hot encoding in the zero-th layer.
A time kernel function developed by Xu {et al.}~\cite{tgat_iclr20} is used to encode the timestamp with cosine functions, defined as
\begin{equation}
  \Phi(\Delta t) = {concat}({cos}(\omega_1 \Delta t), \cdots, {cos}(\omega_d \Delta t)),
  \label{eq:time-encode}
\end{equation}
where $\Delta t = t - t_k$ is the timespan between two timestamps, $\omega_1$ is a trainable parameter of the frequency of the cosine function, and $d$ is the dimension of the output vector.
These dense vectors are concatenated to represent the interaction embedding as
\begin{equation}
    \bm{h}_e^{l-1} = {concat}(\bm{h}_v^{l-1}, \Phi(\Delta t), \bm{m}_{uv}^{t_k}).
    \label{eq:edge-feat}
\end{equation}
For each neighbor set, we employ an attention-based graph convolution layer, written as
\begin{equation}
  \begin{split}
    \alpha_{ue} &= \underset{e \in S(u, t)}{softmax} \{(W_Q \bm{h}_u^{l-1}) (W_K \bm{h}_e^{l-1})^\intercal\}, \\
    \bm{h}_u(t) &= \sum_{e \in S(u, t)} \alpha_{ue} W_V \bm{h}_e^{l-1},\\
  \end{split}
  \label{eq:first-order}
\end{equation}
where $\bm{h}_u^{l-1}(t)$ is the embedding of $u$ at $(l-1)$-th layer, $W_Q, W_K, W_V$ are transform matrices, $\alpha_{ue}$ is the attention score of the interaction $e=(u,v,t_k)$ for node $u$, and the timestamps of nodes are omitted for clarity.

\subsubsection{Sampling Subgraph Fusion}
Let $\bm{h}_u^{TDS}(t)$ and $\bm{h}_u^{GAS}(t)$ be the node embeddings of $u$ with different structure samplers, where the layer of graph convolution is omitted for clarity.
Since TDS and GAS are devised for different purposes, their output node embeddings represent nodes' specific structure preferences.
Therefore, an additional fusion layer is proposed as
\begin{equation}
    \omega_u^S = \bm{q}^\intercal \cdot {sigmoid}(W^S \bm{h}_u^S + b^S), S \in \{\textrm{TDS, GAS}\},
\end{equation}
where $\bm{q}$ is a shared attention vector, $W^S$ is the specific transform matrix for each sampler, and $b$ is the bias term.
The attention scores for different samplers are normalized via a softmax function, written as 
\begin{equation}
    \alpha_u^S =\frac{exp(\omega_u^S)}{exp(\omega_u^{TDS}) + exp(\omega_u^{GAS})}.
    \label{eq:fusion}
\end{equation}
The node embedding of $u$ is finally fused by
\begin{equation}
    \bm{h}_u(t) = \sum_S \alpha^S h_u^S, S \in \{\textrm{TDS, GAS}\}.
\end{equation}

\subsubsection{Optimization for Graph Fusion}
\label{sec:model-opt}

Let $h_u^l(t)$ be the node embedding of $u$ after $l$-layer graph fusion, coupled with our proposed TDS and GAS samplers.
To predict the connection between node pairs, the output probability of the connection is defined as
\begin{equation}
  \hat{y}_{uv}^t = {sigmoid}(W\times{ReLU}(W_u \bm{h}_u^l(t) + W_v \bm{h}_v^l(t))),
  \label{eq:pred}
\end{equation}
where $W, W_u, W_v$ are transform matrices.
The obtained high-order embeddings of $(u, v)$ are fed into Eq.~(\ref{eq:pred}) to get the probability of the edge existence.
The cross-entropy loss is adopted to classify the existence of the edge, which is defined as follows,
\begin{equation}
  \mathcal{L} = - \sum_{(u, v, t) \in E} \{ {log}(\hat{y}_{uv}^t) - c \cdot \mathbb{E}_{j \sim P_n(u)} {log}(\hat{y}_{uj}^t) \},
  \label{eq:loss-ce}
\end{equation}
where $P_n(u)$ is the negative sampling distribution, and $c$ is the number of negative samples. In practice, a uniform distribution over nodes is used as $P_n(u)$, and $c$ is simply set to 1.
The model parameters are then updated using the Adam~\cite{adam} optimizer, which uses the weight-decay strategy to regularize the magnitude of model parameters.

\section{Experiment}
\label{sec:experiment}

\begin{table}[t]
  \caption{Statistics of temporal networks. 
  $\vert V \vert, \vert E \vert$ are the number of nodes and interactions in the network respectively. 
  The graph density is computed by $\frac{\vert V \vert (\vert V \vert - 1)}{2 \vert E \vert}$. 
  The repetition of interactions describes that a node interacts with the same neighbor last time. 
  The time unit of the timespan is one day.}
  \centering
  \resizebox{1.0\linewidth}{!}{
  \begin{tabular}{lrrlrr}
    \toprule
    Temporal Graph             & $\vert V \vert$ & $\vert E \vert$ & Density  & Repetition & Timespan \\
    \midrule
    \multicolumn{6}{c}{Temporal Link Prediction}   \\
    \midrule
    ia-workplace-contacts      & 92              & 9.8K            & 2.34  & 77.1\%    & 11.43           \\
    ia-contacts-hypertext2009    & 113             & 20.8K           & 3.28  & 59.0\%     & 2.46            \\
    ia-contact  & 274             & 28.2K           & 0.75   & 6.9\%       & 3.97            \\
    fb-forum     & 899             & 33.7K           & 0.08  & 20.8\%    & 164.49          \\
    soc-sign-bitcoin  & 5.8K            & 35.5K           & 0.002   & 0.0\%    & 1903.27         \\
    ia-retweet-pol & 19K             & 61.1K           & 0.0003  & 4.7\%   & 48.78           \\
    ia-radoslaw-email      & 167             & 82.9K           & 5.98   & 18.8\%   & 271.19          \\
    soc-wiki-elec      & 7.1K            & 107.0K          & 0.004  & 0.2\%    & 1378.84         \\
    ia-primary-school-proximity     & 242             & 125.7K          & 4.31  & 38.3\%   & 1.35            \\
    ia-slashdot-reply-dir    & 51K             & 140.7K          & 0.0001 & 4.2\%    & 977.36          \\
    \midrule
    \multicolumn{6}{c}{Temporal Node Classification} \\
    \midrule
    Wikipedia & 9.2K   & 157.4K      & 0.0036   & 79.1\%   &  29.77        \\
    Reddit      &  10.9K &    672.4K      & 0.011  & 61.4\%  &   31.00    \\
    \bottomrule
  \end{tabular}
  }
  \label{tab:data}
\end{table}

\subsection{Datasets}
The temporal networks are obtained from Network Repository\footnote{http://networkrepository.com/dynamic.php}~\cite{network-data} and SNAP\footnote{http://snap.stanford.edu/jodie/}~\cite{snapnets,kumar2019jodie}.
Especially, datasets with no node features or edge features collected from Network Repository are used for temporal link prediction.
Datasets with preprocessed edge features collected from SNAP  are used for temporal node classification.
For clarity, in this work, we refer to the network with more than 5,000 nodes as a \textbf{large} network, and the network with less than 1,000 nodes as a \textbf{small} network.
Moreover, the density of a \textbf{dense} network is larger than 1.0, and the density of a \textbf{sparse} network is less than 0.01.
Overall, the variety of datasets could fully demonstrate the effectiveness of compared methods.

\begin{table*}[t]
  \caption{Performance of temporal link prediction over five runs of each method. 
  The symbol '---' is used for failed experiments.
  Bold font indicates the best performance.
  '*' denotes that DPS outperforms the best baseline performance statistically significant with $p < 0.05$ under the two-sided t-test.
  The bottom line denotes the improvement percentages of DPS against the best baseline. 
  }
  \centering
  \resizebox{.9\textwidth}{!}{
  \begin{tabular}{lcccccccccc}
    \toprule
    & \multicolumn{2}{c}{ \textbf{ ia-workplace}} & \multicolumn{2}{c}{ \textbf{ ia-hypertext}} & \multicolumn{2}{c}{ \textbf{ ia-contact }}  & \multicolumn{2}{c}{ \textbf{ fb-forum }} & \multicolumn{2}{c}{ \textbf{soc-bitcoin}}                                                                                                    \\
    \cmidrule(){2-11}
    & Accuracy & AUC  & Accuracy & AUC  & Accuracy & AUC  & Accuracy & AUC  & Accuracy & AUC \\
    \midrule
    Node2Vec\cite{grover2016node2vec} &  0.649 & 0.688  &  0.642 & 0.678  &  0.760 & 0.801  &  0.744 & 0.823  &  0.708 & 0.774 \\
    SAGE\cite{hamilton2017graphsage}  &  0.746 & 0.857  &  0.709 & 0.830  &  0.818 & 0.856  &  0.636 & 0.724  &  0.651 & 0.734 \\
    CTDNE\cite{nguyen2018ctdne} &  0.625 & 0.673  &  0.546 & 0.572  &  0.821 & 0.851  &  0.745 & 0.817  &  0.778 & 0.836 \\
    HTNE\cite{zuo2018htne} &  0.621 & 0.661  &  0.517 & 0.540  &  0.806 & 0.831  &  0.668 & 0.715  &  0.611 & 0.639 \\
    JODIE\cite{kumar2019jodie} &  0.538 & 0.600  &  0.610 & 0.667  &  0.812 & 0.850  &  0.632 & 0.751  &  0.814 & 0.880 \\
    APAN\cite{wang2020apan} &  --- & ---  &  --- & ---  &  --- & ---  &  0.731 & 0.829  &  0.741 & 0.793 \\
    TGAT\cite{tgat_iclr20} &  0.878 & 0.959  &  0.894 & 0.959  &  0.883 & 0.921  &  0.794 & 0.878  &  0.811 & 0.872 \\
    \midrule
    DPS w/ TDS &  0.882 & 0.955  &  0.887 & 0.955  &  0.865 & 0.910  &  0.792 & 0.880  &  0.816 & 0.912 \\
    DPS w/ GAS &  0.914 & 0.975  &  0.910 & 0.969  &  0.895 & 0.928  &  0.798 & 0.885  &  0.821 & 0.913 \\
    DPS w/o Fusion &  0.916 & 0.977  &  0.904 & 0.966  &  0.893 & 0.928  &  $\bm{0.804}$* & $\bm{0.890}$*  &  0.829 & 0.915 \\
    DPS & $ \bm{ 0.918 } $* & $ \bm{ 0.978 } $*  & $ \bm{ 0.911 } $* & $ \bm{ 0.970 } $*  & $ \bm{ 0.896 } $* & $ \bm{ 0.933 } $*  & 0.799 & 0.888 & $ \bm{ 0.837 } $ & $ \bm{ 0.917 } $* \\
    \midrule
    Improvements & 4.6\% & 1.9\% & 1.9\% & 1.1\% & 1.5\% & 1.3\% & 0.6\% & 1.2\% & 2.9\% & 4.1\% \\
    \toprule
    & \multicolumn{2}{c}{ \textbf{ ia-retweet}}   & \multicolumn{2}{c}{ \textbf{ ia-radoslaw}}  & \multicolumn{2}{c}{ \textbf{ soc-wiki}}  & \multicolumn{2}{c}{ \textbf{ ia-primary }} & \multicolumn{2}{c}{ \textbf{ ia-slashdot }}                                                                                                    \\
    \cmidrule(){2-11}
    & Accuracy & AUC  & Accuracy & AUC  & Accuracy & AUC  & Accuracy & AUC  & Accuracy & AUC \\
    \midrule
    Node2Vec\cite{grover2016node2vec} &  0.691 & 0.754  &  0.708 & 0.773  &  0.512 & 0.508  &  0.586 & 0.622  &  0.723 & 0.800 \\
    SAGE\cite{hamilton2017graphsage}  &  0.721 & 0.818  &  0.804 & 0.894  &  0.563 & 0.577  &  0.856 & 0.929  &  0.644 & 0.782 \\
    CTDNE\cite{nguyen2018ctdne} &  0.772 & 0.846  &  0.723 & 0.795  &  0.558 & 0.606  &  0.593 & 0.637  &  0.782 & 0.851 \\
    HTNE\cite{zuo2018htne} &  0.714 & 0.741  &  0.679 & 0.744  &  0.536 & 0.535  &  0.559 & 0.613  &  0.660 & 0.691 \\
    JODIE~\cite{kumar2019jodie} &  0.771 & 0.864  &  0.713 & 0.784  &  0.683 & 0.801  &  0.560 & 0.588  &  --- & --- \\
    APAN\cite{wang2020apan}  &  0.774 & 0.861  &  --- & ---  &  0.719 & 0.775  &  --- & ---  &  0.801 & 0.892 \\
    TGAT~\cite{tgat_iclr20} &  0.813 & 0.889  &  0.830 & 0.905  &  0.819 & 0.869  &  0.910 & 0.960  &  0.611 & 0.604 \\
    \midrule
    DPS w/ TDS &  $\bm{0.851}$* & $\bm{0.928}$*  &  0.803 & 0.887  &  0.761 & 0.901  &  0.907 & 0.952  &  $\bm{0.869}$* & 0.939 \\
    DPS w/ GAS &  0.841 & 0.922  &  0.845 & 0.923  &  0.823 & 0.921  &  0.911 & 0.963  &  0.852 & 0.934 \\
    DPS w/o Fusion &  0.850 & 0.928  &  0.847 & 0.925  &  0.763 & 0.908  &  0.907 & 0.959  &  0.855 & 0.938 \\
    DPS & 0.850 & 0.927 & $ \bm{ 0.859 } $* & $ \bm{ 0.933 } $*  & $ \bm{ 0.835 } $ & $ \bm{ 0.922 } $*  & $ \bm{ 0.928 } $* & $ \bm{ 0.971 } $*  & 0.868 & $ \bm{ 0.944 } $* \\
    \midrule
    Improvements & 4.6\% & 4.3\% & 3.5\% & 3.1\% & 1.9\% & 6.1\% & 1.9\% & 1.1\% & 8.4\% & 5.9\% \\
    \bottomrule
  \end{tabular}
  }
  \label{tab:auc}
\end{table*}

\subsection{Tasks}
\subsubsection{Temporal Link Prediction}
The task is predicting the edge existence given a node pair and a timestamp. 
The datasets are split chronologically that the first 70\% of edges are taken for training, the next 15\% are used for validation, and the final 15\% are used for test evaluation. 
The unseen nodes in the training data are removed due to the lack of node features or edge features.
The observed edges are treated as positive samples, and the negative samples are sampled by substituting the target node of the positive edges with a never-seen node. 
For each dataset, we generate a labeled test set for evaluation.

\subsubsection{Temporal Node Classification}
The task is predicting a user's state that whether the user is banned from the Wikipedia page and subreddit.
The training, validation, and test sets are also split as 70:15:15 for temporal node classification.
The labels of the additional two datasets are extremely imbalanced: {Wikipedia} has 217 positive labels with 157,474 interactions (=0.14\%), while {Reddit} has 366 true labels among 672,447 interactions (=0.05\%). 
The features of user edits in Wikipedia and user posts in Reddit are both converted into the 172-dimensional vectors under the {linguistic inquiry and word count} (LIWC) categories~\cite{liwc2001}.  

\subsubsection{Evaluation Metrics}
For temporal link prediction, we compute the classification \emph{accuracy} and the \emph{area under the ROC curve} (AUC-ROC) when obtaining the prediction probabilities of edge existence.
For temporal node classification, we use the AUC-ROC due to the extremely imbalanced labels.

\subsection{Baselines}
\subsubsection{Temporal Link Prediction}
The baseline methods are initially designed to perform temporal link prediction, categorized as follows:
\begin{itemize}
  \item \textbf{Static graph methods.} Node2Vec~\cite{grover2016node2vec} is an effective skip-gram method for node embedding by exploring the neighborhood with controllable random walks. SAGE~\cite{hamilton2017graphsage} is a strong baseline of graph neural networks, which adopts MaxPooling aggregator for its best performance in the original paper. For static graph methods, node embeddings obtained in the training set are used for temporal link prediction in the test set.
  \item \textbf{Continuous-time graph methods.} CTDNE~\cite{nguyen2018ctdne} generates a final hidden embedding for each node with temporal random walks. HTNE~\cite{zuo2018htne} firstly introduces the Hawkes process in temporal network modeling. JODIE~\cite{kumar2019jodie}, TGAT~\cite{tgat_iclr20}, and APAN~\cite{wang2020apan} are three state-of-the-art temporal network methods. 
\end{itemize}

\subsubsection{Temporal Node Classification}
Continuous-time graph methods predict users' states using temporal node embeddings.
For other methods, we concatenate the node embeddings of the user and the corresponding Wikipedia page or subreddit as input features.
We use the three-layer MLP~\cite{tgat_iclr20} as a classifier, whose hidden dimensions are \{80,10,1\} respectively. 
In addition, the MLP classifier is trained with the Adam optimizer, the Glorot initialization, and the early-stopping strategy with ten epochs. 
Due to the data imbalance, the positive labels are oversampled to achieve the balance label ratio in each batch. 

\subsection{Implementation Details of DPS}
\subsubsection{Temporal Link Prediction}
Our proposed framework is implemented using numpy, cvxpy~\cite{cvxpy,cvxpy2}, and PyTorch~\cite{pytorch2019paszke}.
The TDS sampler optimizes the log-sum-exp objective of Eq.~(\ref{eq:tds-convex}) of each node, which approximates the optimal decay rate by sampling at most 100 terms from a node's temporal interactions.
The GAS sampler optimizes the temporal link prediction objective of Eq.~(\ref{eq:loss-ce}) using a one-layer graph neural network and the Gumbel-softmax trick~\cite{gumbel-softmax}.
The inputs of graph neural networks are concatenated edge features of Eq.~(\ref{eq:edge-feat}), where node features are one-hot encodings if not provided.
The implementation of the attention mechanism is inspired by researchers~\cite{attention2017vaswani,tgat_iclr20}, including the multi-head attention layer of Eq.~(\ref{eq:first-order}) and the fusion layer of Eq.~(\ref{eq:fusion}).
The temporal link prediction of Eq.~(\ref{eq:loss-ce}) is minimized by the Adam optimizer. 
For the \text{hyper-parameters} of model architecture, the number of graph fusion layers is searched over $\{1, 2\}$, the number of attention head is searched over $\{1, 2, 4\}$, and the number of sampled neighbors is searched over $\{10, 20, 30, 40\}$.
For the hyper-parameters of training, the batch size is selected among $\{100, 150, 200, 250\}$, and the dropout ratio is selected among $\{0.0, 0.1, 0.2, 0.3\}$.
We also employ the early-stopping strategy until the validation AUC score does not improve over three epochs. 

\subsubsection{Temporal Node Classification}
Since our DPS can produce temporal node embeddings directly, we use the same three-layer MLP of baseline methods except that the input features are only users' temporal node embeddings.

\subsection{Experiment Comparison}

\subsubsection{Temporal Link Prediction}

\begin{itemize}
  \item Compared with the performance degradation of several temporal network methods, Node2Vec and SAGE are two robust baselines on most datasets.
  \item TGAT performs significantly better than CTDNE and HTNE on most datasets, demonstrating the superiority of graph neural networks. Nevertheless, TGAT shows performance degradation on the large and sparse network, namely \emph{ia-slashdot}. In contrast, APAN performs the second-best on \emph{ia-slashdot} among other methods. However, the implementation of APAN makes it incapable of temporal networks of only a few nodes.
  \item JODIE, as a recurrent evolving method, not only underperforms TGAT on most datasets but also shows a performance decline on four small and dense temporal networks, namely \emph{ia-workplace}, \emph{ia-hypertext}, \emph{fb-forum}, and \emph{ia-primary}. It indicates the recurrent models may be unsuitable for fast changes of user preferences.
  \item 
  Our proposed DPS achieves the best performance of Accuracy and AUC scores on all temporal networks.
  Its three variants also achieve better or comparable performance against the state-of-the-art methods.
  Detailedly, DPS performs much better than the second-best baseline in low-repetition networks, including \emph{soc-bitcoin}, \emph{ia-retweet}, \emph{soc-wiki}, and \emph{ia-slashdot}.
  It is mainly caused by the interaction sparsity of these temporal networks that DPS can capture a node's dynamic preferences from its historical interactions.
  DPS also obtains consistent improvements over baseline methods for other high-repetition networks.
\end{itemize}

\subsubsection{Temporal Node Classification}

\begin{table}[t]
    \caption{AUC scores for temporal node classification over five runs of each method.}
    \centering
    \resizebox{.9\linewidth}{!}{
    \begin{tabular}{lcc}
        \toprule
        Methods & Wikipedia & Reddit \\
        \midrule
        Node2Vec\cite{grover2016node2vec} & $0.812 \pm 0.018$ & $0.618 \pm 0.050$ \\
        SAGE\cite{hamilton2017graphsage} & $0.824 \pm 0.007$ & $0.612 \pm 0.006$ \\
        CTDNE\cite{nguyen2018ctdne} & $0.759 \pm 0.005$ & $0.594 \pm 0.006$ \\
        HTNE\cite{zuo2018htne} & $0.742 \pm 0.011$ & $0.612 \pm 0.009$ \\
        JODIE~\cite{kumar2019jodie} & $0.832 \pm 0.005$ & $0.599 \pm 0.021$ \\
        APAN\cite{wang2020apan}  & $0.899 \pm 0.003$ & $0.653 \pm 0.004$ \\
        TGAT\cite{tgat_iclr20} & $0.837 \pm 0.007$ & $0.656 \pm 0.007$ \\
        DPS & $\bm{0.902 \pm 0.005}$ & $\bm{0.703 \pm 0.004}$* \\
        \bottomrule
    \end{tabular}
    }
    \label{tab:nc}
\end{table}

There are only 44 positive labels in the test set of Wikipedia and 94 positive labels in the test set of Reddit, indicating the user is banned from the Wikipedia page or the subreddit.
Table~\ref{tab:nc} presents the AUC-ROC scores of our DPS and other compared temporal network methods.
Firstly, static graph methods (Node2Vec and SAGE) outperform temporal network methods (CTDNE and HTNE), implying that the banned users have discriminative features. 
Secondly, continuous-time graph methods with GNNs (APAN, TGAT, and DPS) perform much better than other baseline methods, which shows the superiority of GNNs on learning graph structures.
Finally, our DPS performs comparably with APAN on Wikipedia and much better than others on Reddit, validating the robustness of DPS.

\subsection{Ablation Study}

\subsubsection{DPS w/ TDS}
The TDS is designed to capture nodes' temporal patterns by personalized decay rates for each node.
As shown in Table~\ref{tab:auc}, \emph{DPS w/ TDS} beats most baseline methods on all networks.
It even performs better than DPS on two low-repetition networks \emph{ia-retweet} and \emph{ia-slashdot}, which indicates that the objective of Eq.~(\ref{eq:tds-convex}) is effective for sparse networks.
Compared with the robust baseline TGAT~\cite{tgat_iclr20}, the deficiency of TDS on a few networks reveals the importance of node semantics in temporal networks.

\subsubsection{DPS w/ GAS}
The GAS, complementary to the proposed TDS, is designed to capture the semantic proximity among nodes adaptively.
As shown in Table~\ref{tab:auc}, \emph{DPS w/ GAS} beats nearly all the baseline methods on all networks.
It implies that heuristic sampling strategies used in SAGE~\cite{hamilton2017graphsage}, TGAT~\cite{tgat_iclr20}, and APAN~\cite{wang2020apan} may be suboptimal for graph learning of temporal networks.
Moreover, the superiority of DPS to \emph{DPS w/ GAS} shows that the complementary TDS sampler can further boost the performance.

\subsubsection{DPS w/o Fusion}
The attention-based fusion layer is employed to overcome the heterogeneity of TDS and GAS.
Table~\ref{tab:auc} shows that \emph{DPS w/o Fusion} can only achieve comparable performance with the better one between \emph{DPS w/ TDS} and \emph{DPS w/ GAS} on most datasets. 
Specifically, \emph{DPS w/o Fusion} even demonstrates performance degradation on the network \emph{soc-wiki}.
Compared with \emph{DPS w/o Fusion}, DPS achieves consistent performance improvements on most datasets, indicating the importance of fusing the heterogeneity of TDS and GAS.

\subsection{Parameter Sensitivity}

\begin{figure}[t]
    \centering
    \includegraphics[width=.9\linewidth]{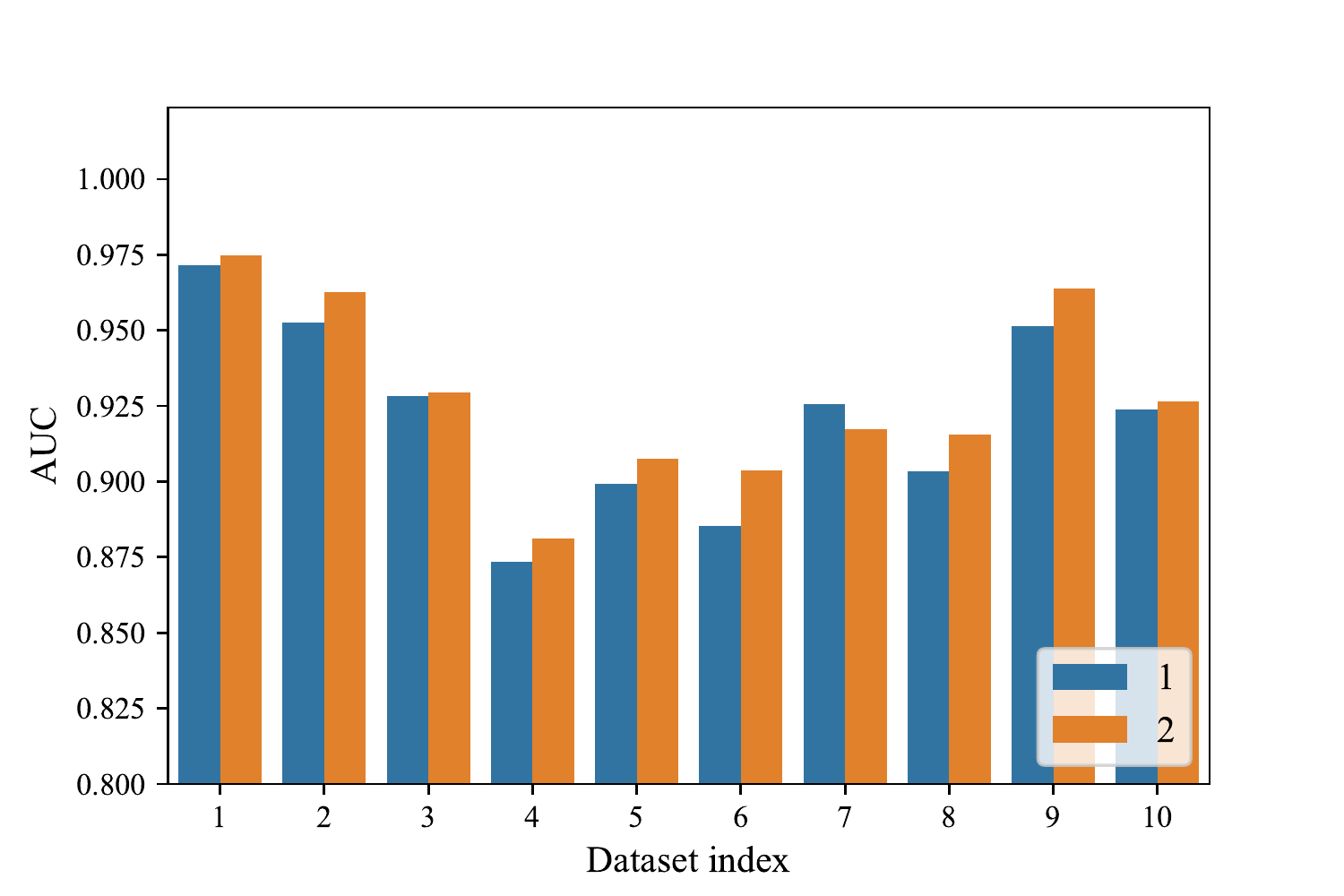}
    \caption{Number of graph fusion layers (over five funs).}
    \label{fig:n_layer}
\end{figure}
\begin{figure}[t]
    \centering
    \includegraphics[width=.9\linewidth]{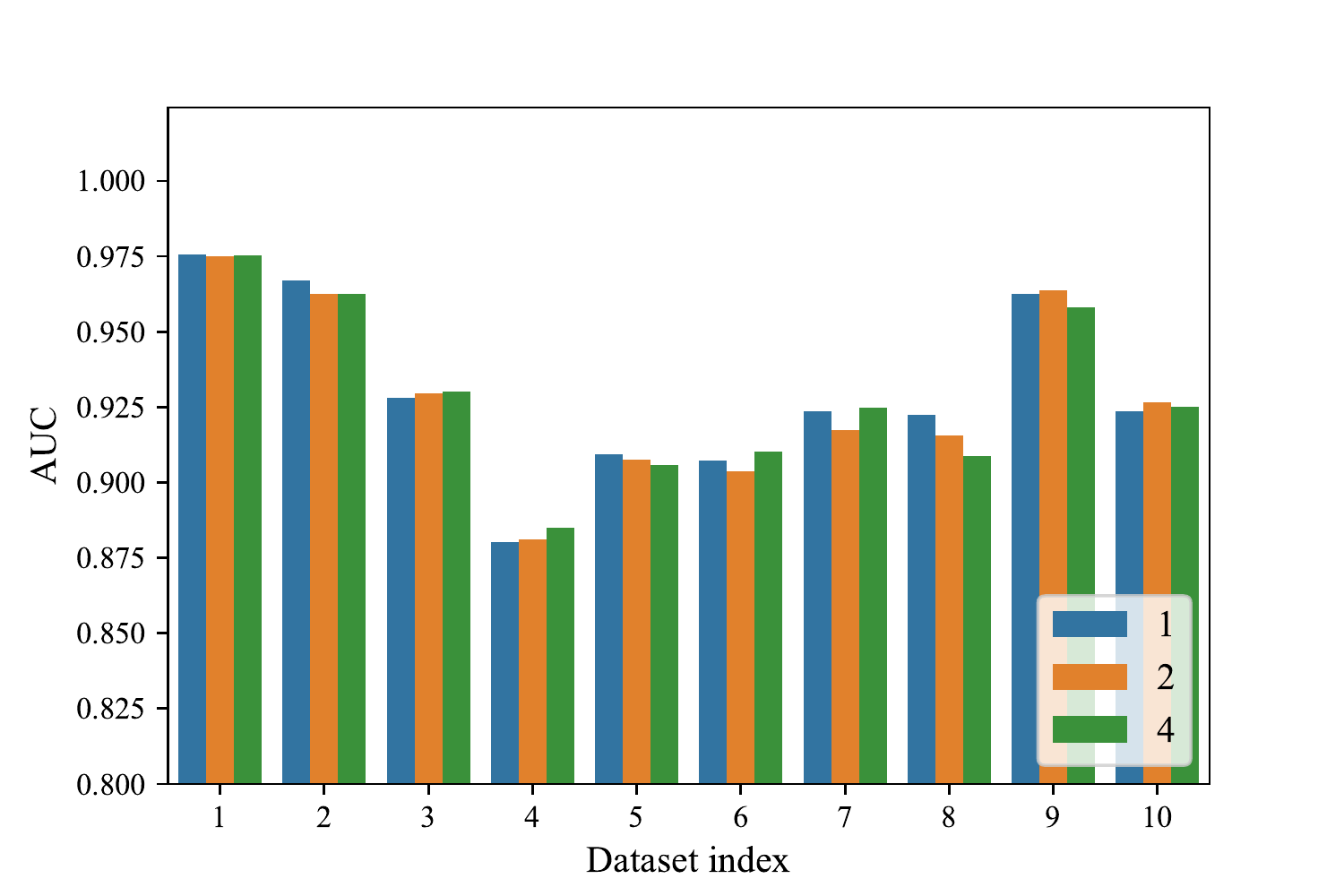}
    \caption{Number of attention heads (over five runs).}
    \label{fig:n_head}
\end{figure}
\begin{figure*}[t]
    \centering
    \includegraphics[width=.9\textwidth]{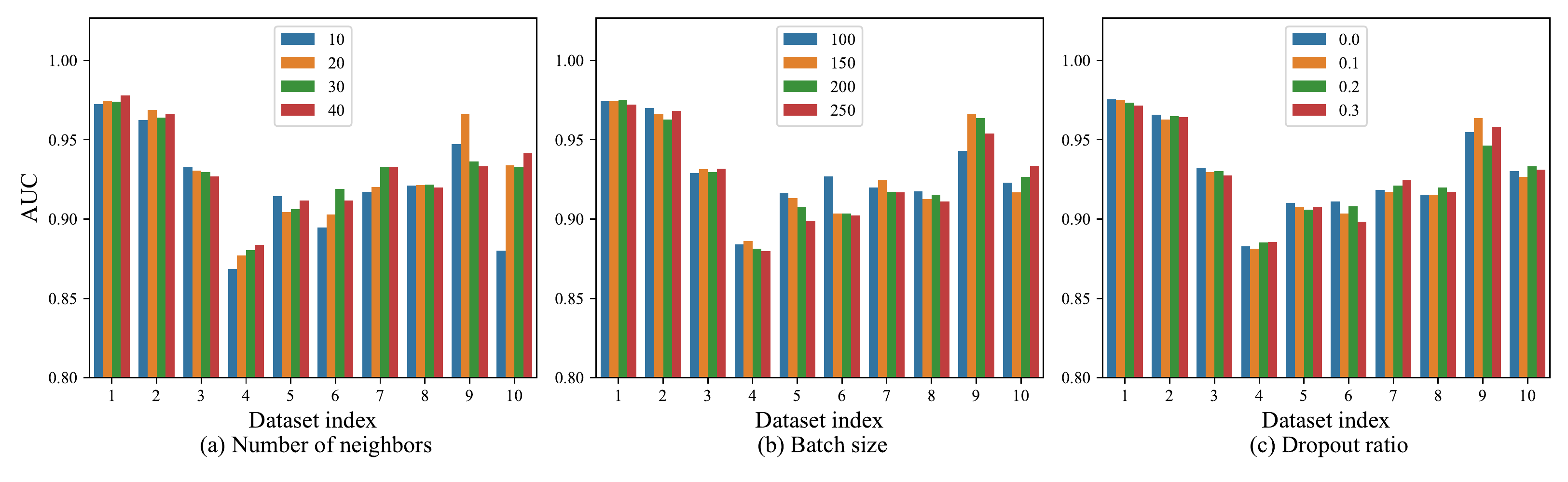}
    \caption{Parameter sensitivity experiments over five runs. Dataset are indexed according to Table~\ref{tab:data}.}
    \label{fig:params}
\end{figure*}

\subsubsection{Number of Graph Fusion Layers}
The number of layers restricts the receptive field of nodes in graph neural networks.
Researchers observed that GNNs first benefit from the increase of layers but later suffer from the over-smoothing and overfitting problem when the number of layers is deep~\cite{deffe2016gcn,kipf2016semi,hamilton2017graphsage,velivckovic2017gat}.
Due to the neighborhood expansion problem, Fig.~\ref{fig:n_layer} only depicts DPS's performance of $\text{layer} =\{1, 2\}$. 
On most datasets, DPS benefits from a deeper layer significantly, which mainly owes to the high-order connectivity~\cite{kipf2016semi,hamilton2017graphsage} and the collaborative filtering signal~\cite{wang2019ngcf}.
In contrast, the performance degradation on the email network \emph{ia-radoslaw} indicates the involved noise of high-order neighbors.

\subsubsection{Number of Attention Heads}
Neural networks employing the attention mechanism~\cite{attention2017vaswani,velivckovic2017gat,tgat_iclr20} often obtain performance improvements with a few attention heads, which enlarges the model capacity of representation learning.
However, our DPS exhibits stable performance regardless of the number of attention heads, which verifies the robustness of our method firstly.
Secondly, our proposed TDS and GAS, which sample the preference structure of networks, actually have the same effect as the attention mechanism, strengthening the preference structure.
Thirdly, the stable performance relieves the burdensome work of tuning hyper-parameters.

\subsubsection{Number of Neighbors}
The number of neighbors is the sampling number of neighborhoods for each node.
Fig.~\ref{fig:params}(a) shows the heterogeneity of different networks that DPS benefits from the increase of neighbors on networks of a long timespan (including \emph{fb-forum}, \emph{ia-retweet}, \emph{ia-radoslaw}, and \emph{ia-slashdot}), while DPS performs stably with different neighbors on networks of a short timespan (including \emph{ia-workplace}, \emph{ia-hypertext}, and \emph{ia-contact}).

\subsubsection{Batch Size}
Deep learning models update the parameters once a mini-batch using the SGD strategy~\cite{alexnet,adam}.
Usually, a smaller batch size accelerates the training speed, and a larger batch size gives a more robust model.
The correlation of performance and batch size drawn in Fig.~\ref{fig:params}(b) can be divided into two kinds: DPS' performance on high-repetition networks such as \emph{ia-workplace}, \emph{ia-hypertext}, \emph{ia-contact}, and \emph{fb-forum} is relatively stable; while a better performance on the rest of networks requires careful hyper-parameter search. 

\subsubsection{Dropout Ratio}
The dropout strategy is a successful technique to avoid overfitting for neural networks~\cite{dropout2012hinton}, which dropouts the neurons randomly and adjusts the hidden outputs accordingly.
Similar to the effects of attention head shown in Fig.~\ref{fig:n_head}, the dropout ratio shown in Fig.~\ref{fig:params}(c) is relatively irrelevant to DPS' performance on most networks.
It may also attribute to the sampling of preference structures, which implicitly regularizes the noise from nodes' neighborhoods.

\section{Conclusion and Future Work}

In this paper, we investigate the dynamic graph sampling problem, recognizing the preference structure of temporal networks adaptively with time elapsing.
Two parameterized samplers, Time Decay Sampling (TDS) and Gumbel Attention Sampling (GAS), are proposed to deal with the personalized temporal patterns of nodes, and temporal node semantics, respectively.
Besides graph neural networks, we devise an attention-based fusion layer for DPS to combine node embeddings from different subgraphs sampled by TDS and GAS.
Extensive experiments validate the rationality and efficiency of learnable structure samplers of temporal networks and embedding fusion of different samplers.

Neighborhood sampling methods~\cite{fastgcn,dropedge2020rong,hamilton2017graphsage} usually aim at boosting training speed and avoiding the over-smoothing problem in graph neural networks.
Moreover, our method also reveals the neighborhood effectiveness for temporal networks.
In the future, we are more interested in developing self-supervised methods that distill the graph structure with several pretraining tasks.
This direction will both improve the task performance and identify the interpretable graph structure for predictions.

\section*{Acknowledgment}

This research was partially supported by the Key-Area Research and Development Program of Guangdong Province (Grant no.2020B0101100005) and Key Research and Development Program of Zhejiang Province (Grant no.2021C01014)

\bibliographystyle{IEEEtran}
\bibliography{ref}

\end{document}